\documentclass[prd,floatfix,superscriptaddress,preprintnumbers,nofootinbib]{revtex4}
\usepackage{float}
\usepackage{epsfig}
\usepackage{epstopdf}
\usepackage{amsmath}
\usepackage{hyperref}
\hypersetup{colorlinks=True,citecolor=blue}
\usepackage{amsfonts}
\usepackage{amssymb}
\usepackage{caption}
\usepackage{float}
\usepackage{bm}
\usepackage{graphicx}
\usepackage{color}
\usepackage{verbatim}
\usepackage{subfigure}

\def\beq{\begin{equation}}
\def\eeq{\end{equation}}
\def\ber{\begin{eqnarray}}
\def\eer{\end{eqnarray}}
\def\benu{\begin{enumerate}}
\def\eenu{\end{enumerate}}

\def\l{\left}
\def\r{\right}

\def\f{\frac}

\def \lleq {\lower0.9ex\hbox{ $\buildrel < \over \sim$} ~}
\def \ggeq {\lower0.9ex\hbox{ $\buildrel > \over \sim$} ~}

\def\prd{{Phys.\@ Rev.\@ D\ }}

\def\plb {{Phys.\@ Lett.\@ B\ }}

\begin{document}
\title{\bf On generality of Starobinsky and Higgs inflation in the Jordan frame}
\author{Swagat S. Mishra}
\email{swagat@iucaa.in}
\affiliation{Inter-University Centre for Astronomy and Astrophysics,
Post Bag 4, Ganeshkhind, Pune 411~007, India}
\author{Daniel M\"uller}
\email{muller@fis.unb.br}
\affiliation{Instituto de F\'{i}sica, Universidade de Bras\'{i}lia, Caixa Postal 04455, 70919-970 Bras\'{i}lia, Brazil}

\author{Aleksey V. Toporensky}
\email{atopor@rambler.ru}
\affiliation{Kazan Federal University, Kazan 420008, Republic of Tatarstan, Russian Federation}
\affiliation{Sternberg Astronomical Institute, Moscow University, Moscow 119991, Russian Federation}

\date{\today}

\begin{abstract}
We revisit the problem of generality of Starobinsky and Higgs inflation. The known results obtained in the Einstein frame
are generalized for the case of an arbitrary initial energy of the scalar field. These results are compared with  the results
obtained directly in the Jordan frame, which, to our knowledge, has not been thoroughly explored in the literature previously.  We demonstrate that the qualitative picture of initial conditions zone 
in the $(\phi, \dot \phi)$ plane, which leads to 
sufficient amount of  inflation,  is quite similar for both the frames in the case of Higgs inflation. For Starobinsky inflation, the conformal
transformation between the  frames relates the geometrical variables in the Jordan frame  with the properties of an effective scalar field in the Einstein
frame. We show that the transformation $(H, R) \to (\phi, \dot \phi)$ is not regular everywhere, leading to some peculiarities in
 the  zone of good initial conditions  in the $(H, R)$ plane.
\end{abstract}


\maketitle
\section{Introduction}
Recent progress in  observations of CMB has led to significant constraints on the viable models of inflation \cite{Akrami:2018odb}.
Very popular models with reasonable physical motivations like the quadratic $\frac{1}{2}m^2\phi^2$ and the quartic  $\lambda \phi^4$ potentials are
now strongly disfavored. The absence of any detection of the primordial tensor modes leads to the conclusion that 
the inflaton potential should  rather be shallow. Other possibilities include the case of $\lambda \phi^4$ potential with the inflaton field 
 non-minimally coupled to the scalar curvature $R$ (this model allows us to consider the Standard Model Higgs field as the inflaton) or the
absence of any fundamental scalar field at all, with inflation induced by the $R^2$ term in the gravitation Lagrangian
(Starobinsky inflation \cite{starobinsky1980new}.)
Note that these two models allow  conformal transformation of the metric into the  Einstein frame
where the action takes the form of  the usual General Relativity action with a minimally coupled  canonical scalar field (in the case of $R^2$ inflation
this field is purely an effective one). The potential required is again a shallow potential.
That is why a detailed analysis of the inflaton dynamics with shallow potentials has  recently become very important. In particular, old classic results of  initial conditions which yield sufficient amount of inflation (that is inflation with at least 60 e-foldings) should be
re-examined for the case of shallow potentials. Recent works on this topic have already indicated certain important
differences from the {\em classic} case of the massive quadratic potential. In particular, starting from  
Planckian initial energy,  initial scalar field values that lie close to $\phi=0$ for the shallow potentials yield, in contrast to the case of the  massive scalar field potential,
sufficient amount of inflation.

The goal of the present paper is to study the generality of inflation while starting from an energy different from the
Planckian one. It has  already been remarked that the requirement of initial Planckian energy for models
with asymptotically flat potentials leads to initial dominance of  the kinetic term. This might be considered to be 
not so natural, if we believe in initial equipartition between the kinetic and the potential term \cite{Gorbunov:2014ewa}.
Starting from  Planckian energy is even less reasonable in the case of  Higgs and Starobinsky inflation -- in both the models considered in the Jordan frame
effective gravitational constant can change in time, and its present value and the value at the early stages of evolution of the Universe
 can differ by orders of magnitude. This means that the  Planck scale, being inversely 
 proportional to the effective gravitational constant, changes during the cosmological evolution
 and there is no reason to take its present value while describing the early Universe.
 We can alternatively use the Einstein frame for description of these inflationary models,
where  the  Planck scale remains constant.
 However, the conformal transformation to the Einstein
frame does not conserve the energy. Hence if we use the Einstein frame as an effective description of Higgs or
Starobinsky inflation (considering the corresponding Jordan frames as the physical ones), initial energy may again
have no connection with the present value of the Planckian energy, despite the fact that
the latter is conserved in the Einstein
frame. 

Quadratic gravity was first addressed by Weyl in 1918 \cite{weyl1918gravitation}, later  reappearing in the 60s and 70s, for example in \cite{buchdahl1962gravitational,Buchdahl:1983zz,tomita1978anisotropic}. After that it has been investigated by many \cite{Berkin:1991nb,Gurovich:1979xg,Cotsakis:1997ck,Cotsakis:2007un,cotsakis2008slice,
Barrow:2005qv,Barrow:2006xb,Barrow:2009gx,Carloni:2007br,Miritzis:2007yn,Miritzis:2003eu}, for a historical review see \cite{schmidt2007fourth}. Starobinsky inflationary model is realized in the particular case of $R^2$ lagrangian \cite{starobinsky1980new}. Also in this connection there is the Ruzmaikina, Ruzmaikin solution \cite{ruzmaikina1970quadratic}. 

On the other hand, in the  context of inflation, analysis  is usually carried out in the  Einstein frame, see for instance \cite{Mishra:2018dtg}. The intention of this present work is to have  specific emphasis on the Jordan frame. This frame is considered as the physical one, and hence, the results obtained in the Jordan frame have more direct physical meaning. Moreover, if  quadratic gravity could be considered as a low-energy approximation of some more general theory ( for a particular example of such theory, see \cite{Koshelev}), the initial conditions good for inflation should be attractors for dynamics in such a full theory
which can be used to check its consistency. Since there is no {\it a priori } reason for 
this underlying general theory to have an Einstein frame description, it is important that the results for initial conditions to be presented in the Jordan frame.
In the present paper we consider initial conditions leading to successful 
Starobinsky inflation (we choose 60 e-folds
as a criterion for inflation to be successful) in the  Einstein frame in section \ref{sec:staro_einstein} as well as in the Jordan frame in section \ref{sec:staro_jordan} before describing the correspondence
between these two frames  in section \ref{sec:staro_jordanvseinstein}. Since the scalar field potential in the Einstein frame for Staroninsky inflation shares some common features with
the scalar field potential in the Einstein frame for the Higgs inflation, it is reasonable to consider  the same
problem for the Higgs inflation also.  We then discuss similarities and differences in the results obtained for these
two popular models\footnote{For investigations in $f(R)$ gravity, see \cite{Elizalde:2018rmz,Alho:2016gzi}}.  Our results are concluded in section \ref{sec:conclusion}.

\section{Starobinsky inflation in the Einstein frame}
\label{sec:staro_einstein}

In the following we will be studying quadratic gravity given by the
action 

\begin{equation}
S_{J}=\int d^4 x \sqrt{-g}\frac{m_p^2}{2}\left[R+\beta R^{2}\right]~,\label{lagrangeana}
\end{equation}
where $m_p=\frac{1}{\sqrt{8\pi G}}$ is the reduced Planck mass.
Note  that (\ref{lagrangeana}) is written in the  Jordan frame which is usually considered to be the physical frame with metric $g_{ab}$, being used to obtain spatial distances and time lapses. On the other hand, following Barrow and Cotsakis \cite{Barrow:1988xh}, the Einstein frame metric is  conformally related to the Jordan frame metric as $\tilde{g}_{ab}=e^{\tilde{\phi}} g_{ab}=f^\prime g_{ab}$. The conformal factor $f^\prime$ stands for the derivative of the particular $f(R)$ with respect to the argument. While the Lagrangian given in (\ref{lagrangeana}) reduces in the isotropic cases to the corresponding $f(R)=R+\beta R^2$, $R$ being the Ricci scalar.

The intention is to analyse the initial conditions for inflation for zero spatial curvature case in the Jordan frame and compare the results obtained
with those in the Einstein frame which have been found before
(for instance  in \cite{Mishra:2018dtg}). Action (\ref{lagrangeana}) in the Einstein frame is given by
\beq
 S_{E}=\int d^4 x \sqrt{-\tilde{g}}\,\frac{m_p^2}{2}\left[\tilde{R}-\frac{3}{2}\tilde{g}^{ab}\partial_a\tilde{\phi}\partial_b\tilde{\phi}-\tilde{V}(\tilde{\phi})\right]~.
 \label{eq:action_star_einstein}
\eeq
We have 
\beq
R=\frac{e^{\tilde{\phi}}-1}{2\beta}~,
\label{RC}
\eeq
and the potential depends on the particular type of $f(R)$ 
\begin{equation}
\tilde{V}(\tilde{\phi})=\frac{Rf^\prime-f}{(f^\prime)^2}~,
\end{equation}
which, for $f(R)=R+\beta R^2$, becomes 

\beq
\tilde{V}(\tilde{\phi})=\f{\left(1-e^{-\tilde{\phi}}\right)^2}{4\beta}~.
\eeq

\begin{figure}[htb]
\begin{center}
\includegraphics[width=0.85\textwidth]{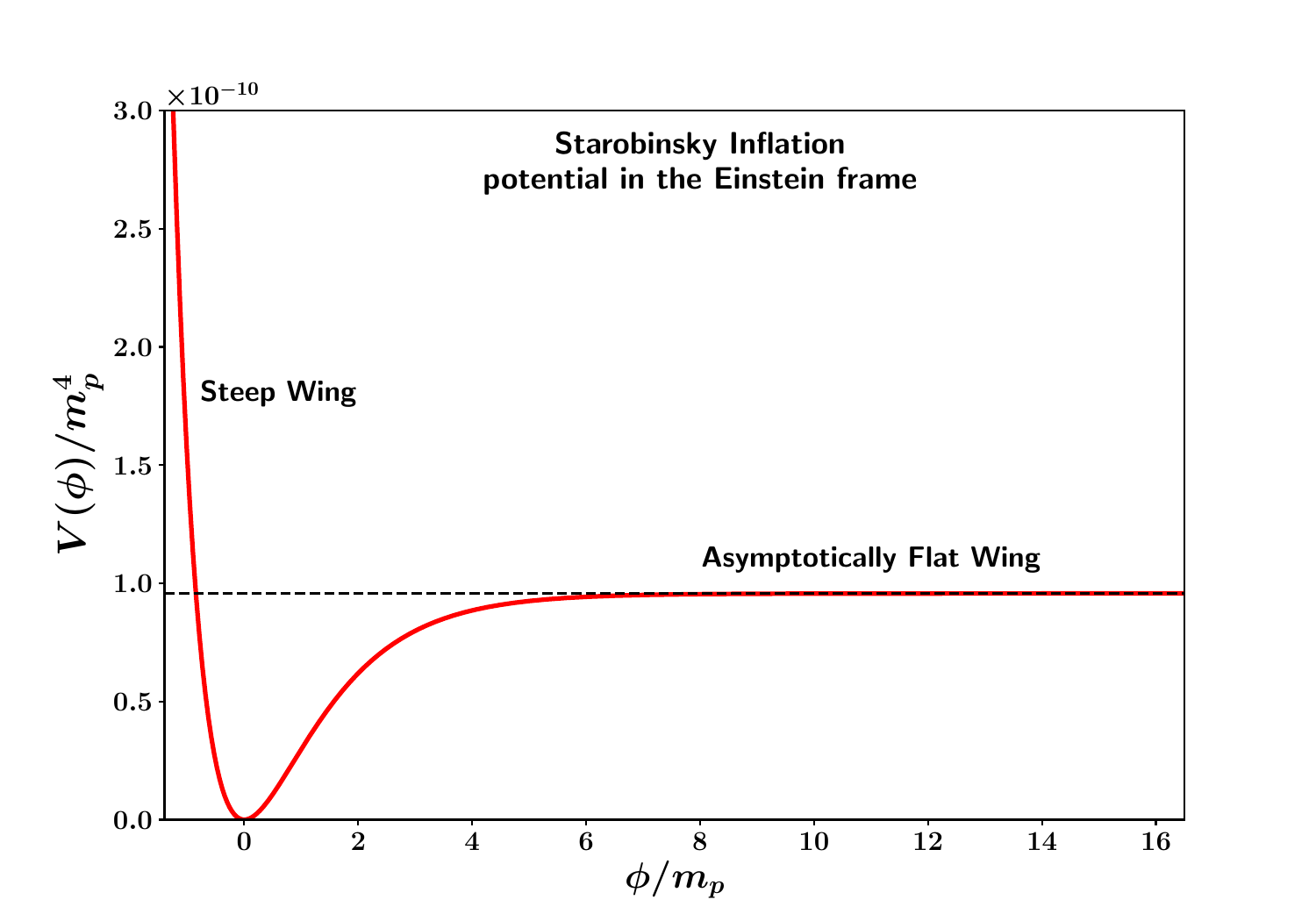}
\captionsetup{
	justification=raggedright,
	singlelinecheck=false
}
\caption{Starobinsky inflation potential in the Einstein frame (\ref{potential}) is shown in this figure to possess an asymptotically flat right wing and a steep left wing.}	
\label{fig:pot_star}
\end{center}
\end{figure}

We also  follow the same units chosen in \cite{Mishra:2018dtg}, in order to obtain the  canonical kinetic term in (\ref{Einstein_frame}) as well as to keep the appropriate dimension of the scalar field by choosing 
\begin{equation}
\tilde{\phi}=\frac{2}{\sqrt{6}}\f{\phi}{m_p}~,\label{phi-phi_tilda}
\end{equation}
and potential 
\begin{equation}
V(\phi)=\frac{m_p^2}{2} \frac{1}{4\beta}\left[1-\exp\left(-\frac{2}{\sqrt{6}}\f{\phi}{m_p}\right)\right]^2~. \label{potential_0}
\end{equation}
By expressing $\beta$ in terms of the mass of the scalaron given by $m=1/\sqrt{6\beta}$, the potential takes the form

\beq
V(\phi)=\frac{3}{4}m^2 m_p^2\left[1-\exp\left(-\frac{2}{\sqrt{6}}\f{\phi}{m_p}\right)\right]^2~,
\label{potential}
\eeq
which is shown in figure \ref{fig:pot_star} to possess an asymptotically flat right wing suitable for inflation. The action in the Einstein frame takes the familiar form 
\begin{equation}
 S_{E}=\int d^4 x \sqrt{-\tilde{g}}\left[\frac{m_p^2}{2}\tilde{R}-\frac{1}{2}\tilde{g}^{ab}\partial_a\phi\partial_b\phi-V(\phi)\right]~, 
 \label{Einstein_frame}
\end{equation}
 which coincides with the action given by equation (61) of \cite{Mishra:2018dtg} with $V(\phi)$ given by (\ref{potential}).
 
Analysis  of generality of inflation for the action (\ref{Einstein_frame}), thoroughly  carried out in
\cite{Mishra:2018dtg}, leads to results rather different from  the already known old results for a massive
scalar field with $\f{1}{2}m^2\phi^2$ potential. Initial conditions were chosen by keeping the initial energy of the scalar field
to be  of the order of Planckian energy, so that initial value of the scalar field $\phi$ fixes the initial value
of $\dot \phi$ up to its sign. The central result of \cite{Mishra:2018dtg} regarding the action (\ref{Einstein_frame}) is that inflation is sufficient (with 60 or higher number
of e-folds), if the initial $\phi\geq 15\, m_p$ for both possible signs of $\dot \phi$. More interesting is the fact that sufficient
inflation can be achieved even  starting from  $\phi\geq -4 \, m_p$ if we fix initial $\dot \phi$ to be positive. Note that it implies starting from initial $\phi=0$ with positive $\dot \phi$, we get enough inflation -- a feature 
clearly different from the massive scalar field case.

Qualitatively this occurs because starting from $\phi=0$ the scalar field has enough kinetic energy to 
climb the shallow right wing of the potential in figure \ref{fig:pot_star}. This also means  that if we consider similar potential,
but possessing both  left and right wings (this is the potential for Higgs inflation in the Einstein frame as shown in figure \ref{fig:pot_higgs}), then 
sufficient inflation can occur for any initial $\phi$, the only restriction being for
$4 \, m_p \leq |\phi|\leq 15 \, m_p $, we would need $\mbox{sgn} (\dot \phi)=\mbox{sgn} (\phi)$, while for $\phi$ outside this range
the sign of initial $\dot \phi$ does not matter  (see figure 16 of \cite{Mishra:2018dtg})\footnote{Please note that we are quoting the field values  $\chi_A$ and $\chi_B$ described in \cite{Mishra:2018dtg} starting from Planckian initial energy.}.

 Such striking difference
between the power-law and the asymptotically flat potentials requires qualitative explanation. Since a symmetric potential is easier
to study and it has its own significance being the Einstein frame potential for the Higgs inflation, we start our
consideration from the symmetric potential before resuming  the study of  Starobinsky inflation in the Section \ref{sec:staro_jordan}.

\section*{{\large Generality of Higgs inflation revisited}}

The action for a scalar field $\varphi$ which couples non-minimally to 
gravity  is given by 
\cite{higgs1,higgs2,kaiser,Mishra:2018dtg} 
\begin{equation}
S_{J}=\int d^{4}x\sqrt{-g} \left[ f(\varphi)R-\frac{1}{2} g^{ab}\partial_{a}\varphi\partial_{b}\varphi-U(\varphi) \right]
\label{eqn:action_higgs}
\end{equation} 
where $R$ is the Ricci scalar and $g_{ab}$ is the  metric in the Jordan frame. 
The potential for the $SM$ Higgs field is given by            
\begin{equation}
U(\varphi)=\frac{\lambda}{4}\left( \varphi^{2}-\sigma^2 \right )^{2}
\label{eqn:higgs_pot1}
\end{equation}
where $\sigma$ is  the vacuum expectation value of the Higgs field 
\begin{equation}
\sigma=246 \,GeV =1.1\times 10^{-16} m_{p}
\label{eqn:higgs_vev}
\end{equation}  
and the Higgs self-coupling constant has the value 
$\lambda=0.1$.
Furthermore      
\begin{equation}
f(\varphi)=\frac{1}{2}(m^{2}+\xi \varphi^{2})
\label{eqn:fun_nm1}
\end{equation}
where $m$ is a mass parameter given by \cite{kaiser}
$$m^{2}=m_{p}^{2}-\xi\sigma^2$$
$\xi$ being the non-minimal coupling constant whose value
\begin{equation}
\xi=1.62\times10^{4}
\label{eqn:coup_nm}
\end{equation}
agrees with CMB observations \cite{Akrami:2018odb,Mishra:2018dtg}.
For the above values\footnote{Note that since the observed vacuum expectation value of the Higgs field $\sigma=1.1\times 10^{-16} m_{p}$ is much smaller compared to the energy scale of inflation we have neglected it from our subsequent calculations.} of $\sigma$ and $\xi$, one finds $m\simeq m_{p}$, so that 
\begin{equation}
f(\varphi) \simeq \frac{1}{2}(m_{p}^{2}+\xi \varphi^{2})=\frac{m_{p}^{2}}{2}\left( 1+\frac{\xi\varphi^{2}}{m_{p}^{2}} \right)~.
\label{eqn:fun_nm2}
\end{equation}
 We now transfer to the Einstein frame by means of the following  conformal transformation of the metric \cite{Mishra:2018dtg}
\begin{equation}
g_{ab}\longrightarrow \tilde{g}_{ab}=\Omega^{2} g_{ab}
\label{eqn:confo_trans}
\end{equation}
where the conformal factor is given by
\begin{equation}
\Omega^{2}=\frac{2}{m_{p}^{2}} f(\varphi)=1+\frac{\xi\varphi^{2}}{m_p^{2}}~.
\label{eqn:confo_fact}
\end{equation}
 After the field redefinition $\varphi\longrightarrow \phi$  the action in the {\em Einstein frame}
 is given by \cite{Mishra:2018dtg}
\begin{equation}
S_{E}=\int d^{4}x\sqrt{-\tilde{g}} \bigg [ \frac{m_{p}^{2}}{2}\tilde{R}-\frac{1}{2} \tilde{g}^{ab}\partial_{a}\phi\partial_{b}\phi-V(\phi) \bigg ]
\label{eqn:action_einstein}
\end{equation}
where 
\begin{equation}
V(\phi)=\frac{U[\varphi(\phi)]}{\Omega^{4}}
\label{eqn:pot1}
\end{equation}
and 
\begin{equation}
\frac{\partial\phi}{\partial\varphi}=\pm\frac{1}{\Omega^{2}}\sqrt{\Omega^{2}+\frac{6\xi^{2}\varphi^{2}}{m_{p}^{2}}}~.
\label{eqn:pot2}
\end{equation}
Eq. (\ref{eqn:action_einstein}) describes General Relativity (GR) in the presence of a 
minimally coupled
scalar field $\phi$ with the potential $V(\phi)$.
(The full derivation of the action in the Einstein frame is  given in  \cite{Mishra:2018dtg,kaiser}.)

\begin{figure}[htb]
\begin{center}
\includegraphics[width=0.85\textwidth]{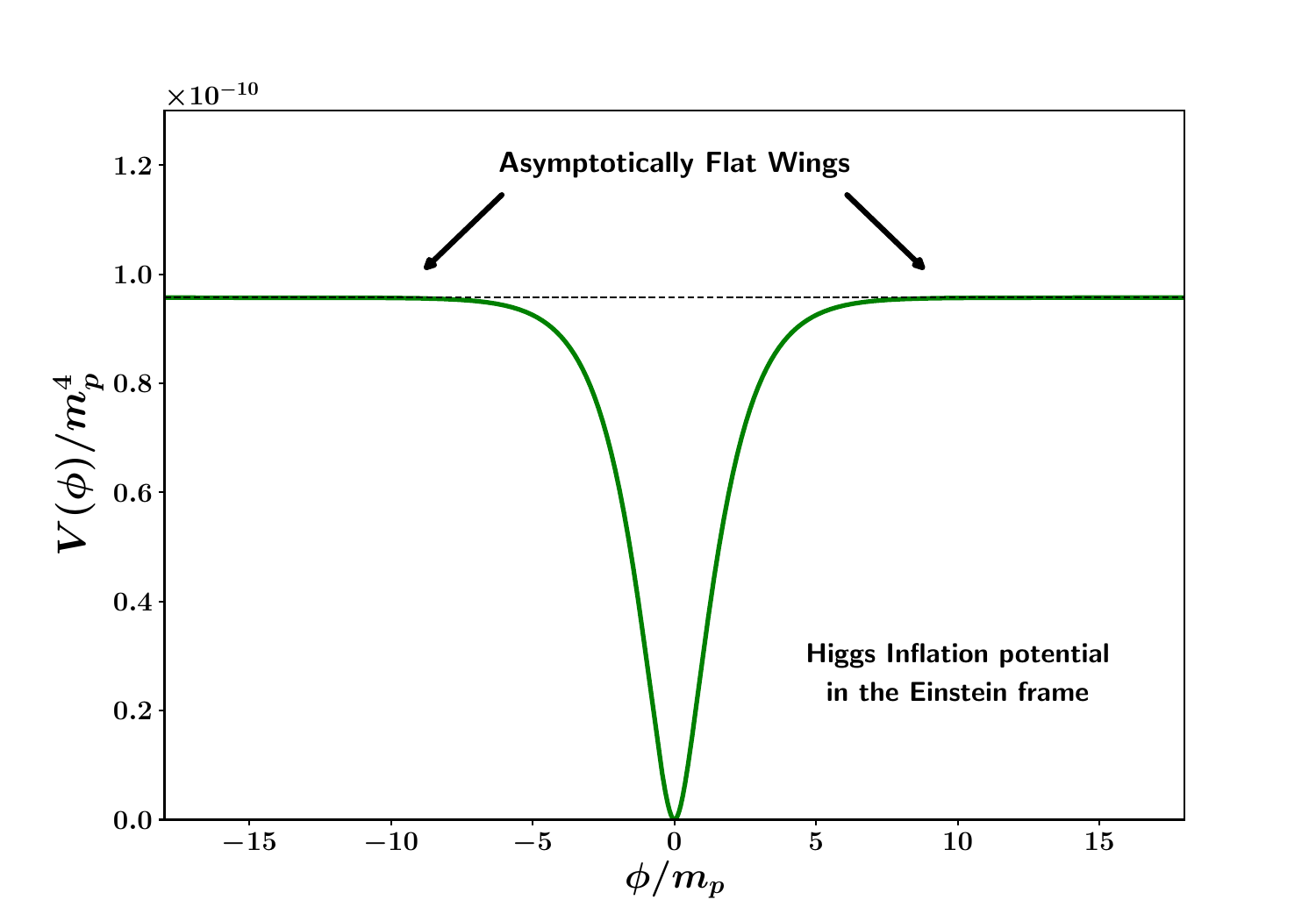}
\captionsetup{
	justification=raggedright,
	singlelinecheck=false
}
\caption{The symmetric Higgs inflation potential in the Einstein frame (\ref{pot}) is shown in this figure to possess two asymptotically flat wings.}	
\label{fig:pot_higgs}
\end{center}
\end{figure}

While considering Higgs inflation in the Einstein frame, there is an approximate analytical form of the  potential (\ref{eqn:pot1}) given in \cite{Mishra:2018dtg} which we reproduce here \begin{equation}
V(\phi)\simeq V_0 \left[1-\exp\left(-\frac{2}{\sqrt{6}}\frac{|\phi|}{m_p}\right) \right]^2~,
\label{pot}
\end{equation}
where $V_0=\lambda m_p^4/(4\xi^2)$. The symmetric potential  potential (\ref{pot}) is shown in figure \ref{fig:pot_higgs} to possess two asymptotically flat wings. We know from figure 16 in \cite{Mishra:2018dtg} that the region near $\phi=0$ gives adequate amount of inflation in contrast to the massive scalar field case.

\subsection{The worst initial condition for inflation}
We start our consideration by noting 
that, contrary to the wide-spread opinion, the $\phi=0$ initial condition is not the worst for inflation
even for a massive scalar field case. 
To support this view let us consider the number of e-folds when starting from Planck energy for different
initial $\phi$ (see figure \ref{fig:chaoticvsflat} below) with the velocity of the scalar field is directed downward i.e $\mbox{sgn} (\dot \phi)=-\mbox{sgn} (\phi)$. We can see clearly that $\phi=0$ initial condition is not the
worst for inflation, though it indeed gives us insufficient number of e-foldings of  inflation for $V(\phi)=\frac{1}{2}m^2 \phi^2$ case. However,
zero inflation corresponds to some finite value of initial $\phi$ at Planck scale given by $\phi_i\simeq 11~ m_p$. Moreover, this value is roughly the same both for the  massive  and  the asymptotically flat (\ref{pot}) potentials. The goal of the present subsection is to find this value analytically.

\begin{figure}[htb]
\begin{center}
\includegraphics[width=0.8\textwidth]{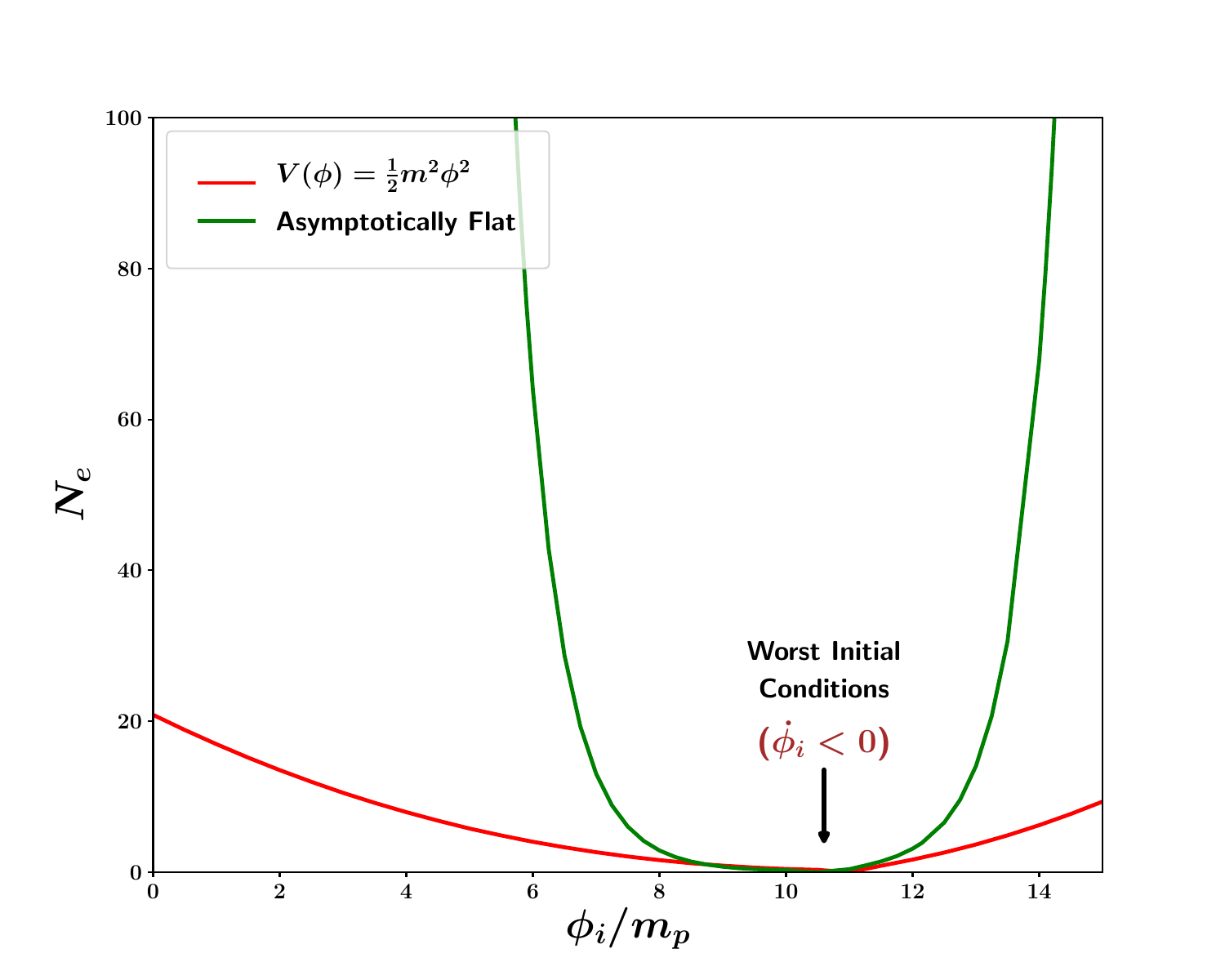}
\captionsetup{
	justification=raggedright,
	singlelinecheck=false
}
\caption{The number of e-foldings obtained before the end of inflation, estimated numerically,   starting from the Planckian initial energy is plotted as a function of the initial field value $\phi_i$ with $\dot \phi_i<0$ for the case of  massive quadratic potential $V(\phi)=\f{1}{2}m^2\phi^2$ in red color and the asymptotically flat Higgs potential in the Einstein frame, given by (\ref{pot}), in green color.  From this figure, it is clear that the worst initial conditions,  leading to no inflation at all, lie near $\phi_i\simeq 11\, m_p$ for both type of potentials. The only difference is that if we start at  $\phi_i=0$, the asymptotically flat potential leads to sufficient amount of inflation while the quadratic potential does not (which was the central result of \cite{Mishra:2018dtg}).}	
\label{fig:chaoticvsflat}
\end{center}
\end{figure}

To do this we remind the  reader another situation  different from our problem here which, however, as we
will show below, can be used to achieve our goal. Namely, consider the hypothetical case of contracting regime of
the Universe filled with a scalar field. It is known that the equation of state $w_{\phi}$ has a different asymptote during contraction as compared to  expansion. In particular, if the potential of the scalar field is less steep than
the exponential one, the field is effectively massless during contraction for almost all
initial conditions fixed at the beginning of the contraction stage \cite{FOSTER}. This means that the effective
equation of state is $w_{\phi}\simeq 1$, scale factor grows with time as $a \sim (t-t_0)^{1/3}$, $H=1/3(t-t_0)$, 
and the scalar field finally reaches the value 
$$
\phi_b=\frac{m_p}{12\pi} \ln{\frac{H_b}{H_{in}}},
$$
where $H_{in}$ and $H_b$ are the Hubble parameters at the commencement and end of this regime $w_{\phi} \simeq 1$ after which universe starts expanding. For 
a massive scalar field, assuming that this
regime ends at the Planck energy and remembering that it starts when scalar field is of the order of $m_p$
(for lower $\phi_i$ we have an amplifying oscillations on contraction stage),
this formula can be rewritten as (see \cite{Sahni:2012er} for details)
\begin{equation}
\phi_b=\frac{m_p}{12\pi} \ln{\frac{m_p}{m}}. \label{v.crit}
\end{equation}

Note, that this formula is also applicable  for a wider class of potentials. Namely, if the potential can be
approximated as $m^2 \phi^2/2$ for $\phi \leq m_p$, so that $w_{\phi} \approx 1$ regime starts when massive
potential approximation is still valid, then scalar field will not feel the behaviour of its
potential from $\phi \approx m_p$ till the Planck energy, because in this regime the potential is
irrelevant. So that we can apply this formula for asymptotically flat potential of the Higgs
inflation. Since this potential has the same  effective mass in the small field limit  as the observationally motivated $m$ for the massive scalar field potential, both the potentials lead
to practically the same $\phi_b$ at the Planck boundary which is about $11\, m_p$.

Now we return to initial conditions which are the worst for inflation.
 To find an analytical answer let us consider
the case of a contracting universe. A typical contracting universe  has an equation of state of a stiff matter i.e $w_{\phi}\simeq1$ as an attractor solution with the field growing fast while  universe approaches the singularity.  Now imagine that we stop the evolution at some energy level $H_b$ and reverse the direction of time. At this point, the field has climbed up to a value $\phi_{b}$ and has a velocity  $\dot{\phi}_b$ of the order of the total energy, since the potential is negligible with respect to the kinetic term. After time reversal $\dot \phi$ changes its sign, and Universe will follow the same trajectory while expanding. Obviously, the equation of state is invariant under time reversal, so the universe will expand with no inflation at all. Hence the  initial conditions $\phi=\phi_{b}$ and  $\dot \phi=-\dot \phi_b$  are  the worst initial conditions for inflation to occur. If we consider Planck
boundary, $\dot \phi^2/2 $ will be the Planck energy (since kinetic term dominates in this regime) and
$\phi$ should be equal to $\phi_b$ which is found earlier. This coincides  with the numerically obtained value of $\phi_i$ presented in figure \ref{fig:chaoticvsflat}.
 As it is known, $w=1$ at expansion (in contrast to contraction) requires a set of  very special  conditions, so for $\dot{\phi_i}$ different enough from the one constructed above,  inflation becomes possible. These qualitative properties are true for both power-law as well as asymptotically flat potentials. 
However, for a massive scalar field on a Planckian boundary, $\phi=\phi_b=11 \, m_p$ is  not high enough to cause sufficient amount of inflation (i.e inflation with  at least $60$ e-foldings) even if we start with $\dot{\phi_i}=0$. That is why the fact that
the worst initial situation occurs for non-zero initial $\phi$ has a little effect for massive potential -- this value is still close to zero  from the viewpoint of adequate inflation.
 On the contrary, this value $\phi=\phi_b$ is high enough for 
successful inflation in the case of plateau potentials (\ref{potential}) and (\ref{pot}). However, large negative $\dot{\phi_i}$
kills the inflation completely, despite of the fact that initial $\phi$ is enough to get 60 e-folds when starting from
$\dot{\phi_i}=0$. That is why the sign of initial $\dot \phi$ is more important for asymptotically flat potentials.

 It is also worth mentioning that we can start from $\phi=0$ and with a Planckian kinetic energy in the massive scalar
field case and get some number of e-foldings of inflation. Numerical simulations show that for $m\simeq 6\times 10^{-6}~m_p$, the field climbs till $\phi_{max}=10.4 ~m_p$ yielding about $28$ e-foldings. This is still not enough, however, it shows the effect of initial kinetic term in   helping the field climb up the potential causing the universe to inflate  even when starting from $\phi=0$.  The same situation happens for an asymptotically flat potential and in this case, this effect is important for realistic models,  because for a such potential, we can get adequate inflation  with $N_e\geq 60$ starting from $\phi=0$.

Summarizing, in both the cases (massive and plateau potentials) initial $\phi=0$ leads to some e-foldings of inflation. While for plateau potentials, we do get sufficient amount of inflation, for a massive scalar field
with the value of the scalar field mass that agrees with observation,  inflation is not adequate. We have
also checked numerically that
for initial energy scale $H$  smaller than $10^{-3}~m_p$ we do not get adequate inflation from $\phi=0$
even for the case of plateau potentials (\ref{potential}) and (\ref{pot}), so that
the structure of initial conditions zone good for inflation becomes similar to such zone for a massive scalar 
field. This means that the structure of the zone of ``good" initial conditions  for plateau potentials is rather 
sensitive to initial energy.

On the other hand, strictly zero
inflation initial conditions necessarily require   non-zero  initial $\phi$.
However, despite common general dynamical features, the set of initial conditions which leads
to more than 60 e-folds starting from the Planck boundary looks quite different in the power-law potentials as compared to  plateau potentials,
as it have been shown in \cite{Mishra:2018dtg}.

\subsection{Initial conditions with arbitrary initial energy scale}

\begin{figure}[htb]
\centering
\subfigure[]{
\includegraphics[width=0.48\textwidth]{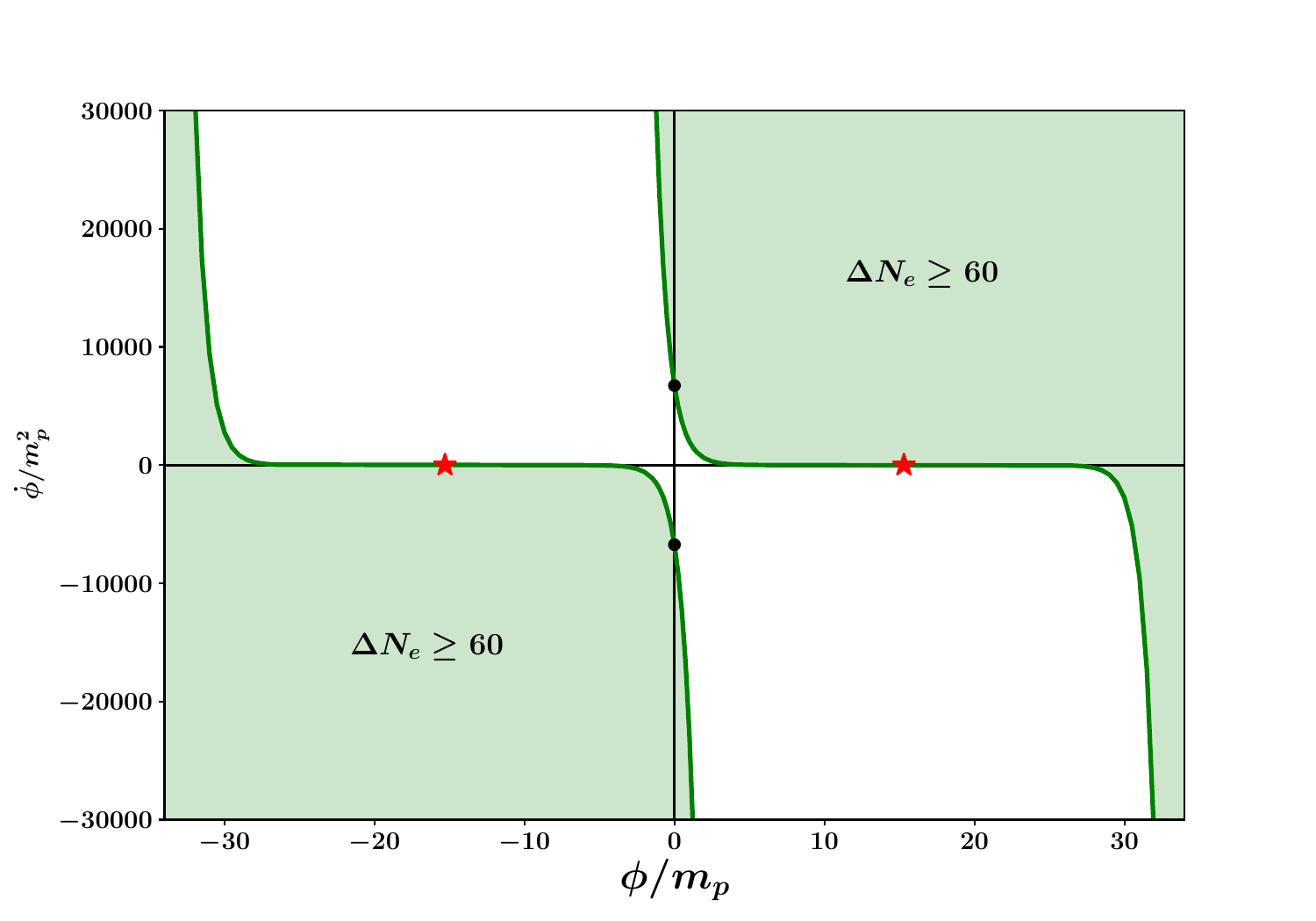}}
\subfigure[]{
\includegraphics[width=0.48\textwidth]{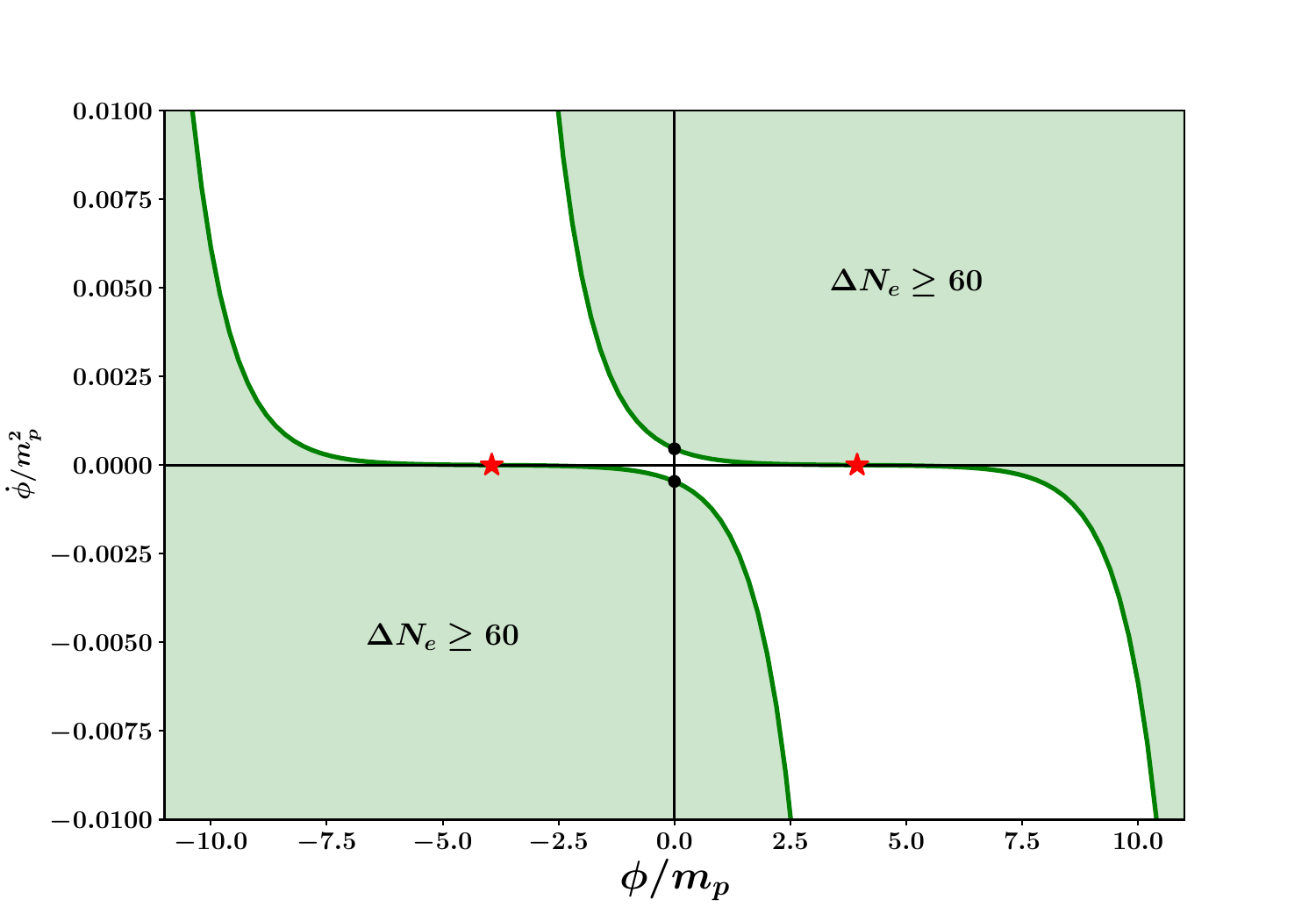}}
\captionsetup{
	justification=raggedright,
	singlelinecheck=false
}
\caption{Above figure represents the basin of initial conditions for  {\bf (a)} $V(\phi)=\f{1}{2}m^2\phi^2$ potential and {\bf (b)}  the Higgs inflation potential in the Einstein frame (\ref{pot}). The green regions contain initial conditions leading to sufficient amount of inflation with at least 60 e-foldings. The initial energy needed can be roughly  characterized by black and red points in the plots -- the black points indicate the least possible value of initial $\dot \phi$ required for sufficient inflation to occur starting from initial $\phi=0$, while the red points indicate the least initial value of  $\phi$ for sufficient inflation to occur starting from initial $\dot \phi=0$.} 
\label{f.swag}
\end{figure}

The results of the previous subsection have been obtained for an initial energy of the order of Planck scale. They become even more clear if we
construct initial conditions leading to at least 60 e-folds in the ($\phi$, $\dot \phi$) plane without fixing
the initial energy scale $\rho_{\phi}$. Our results are presented in figure \ref{f.swag} for the  massive $\f{1}{2}m^2\phi^2$ potential in the left panel and for the asymptotically flat  Higgs potential (\ref{pot}) in the Einstein frame. We can see that the results are qualitatively the same  -- the initial conditions 
located within the white bands in figure \ref{f.swag} $(a)$ and figure \ref{f.swag} $(b)$
do not lead to successful inflation. The only difference between these two figures is the width of this band,
 and this difference is responsible
for the above mentioned fact that initial $\phi=0$ yields  enough inflation  in the  case of plateau-like potentials,
but not in the case of power-law potentials. Hence this difference appears, not because of  some fundamental mathematical features of
inflationary dynamics in these two cases, but rather due to  the particular physical restrictions -- starting
within Planck energy ($\rho_{\phi}=m_p^4$) and demanding at least 60 e-folds.

It should be noted that starting from Planckian energy is a reasonable strategy if we consider the Einstein
frame as a fundamental one. If, on the contrary, we think  the Jordan frame to be  the physical one, then
taking into account that energy is not an invariant of the conformal transformation between the frames, there is
no reason to fix the initial energy in the Einstein frame.
From this point of view the results presented in
figure \ref{f.swag} {\bf (b)}, where we did not fix the initial energy, are more relevant to the problem of generality of the Higgs inflation in the Einstein frame.

\subsection{Higgs inflation in the Jordan frame}
We work directly in the Jordan frame, considering the equations of motion \cite{Sami:2012uh} obtained  from the  action (\ref{eqn:action_higgs}), given by

\begin{figure}[htb]
\centering
\subfigure[]{
\includegraphics[width=0.48\textwidth]{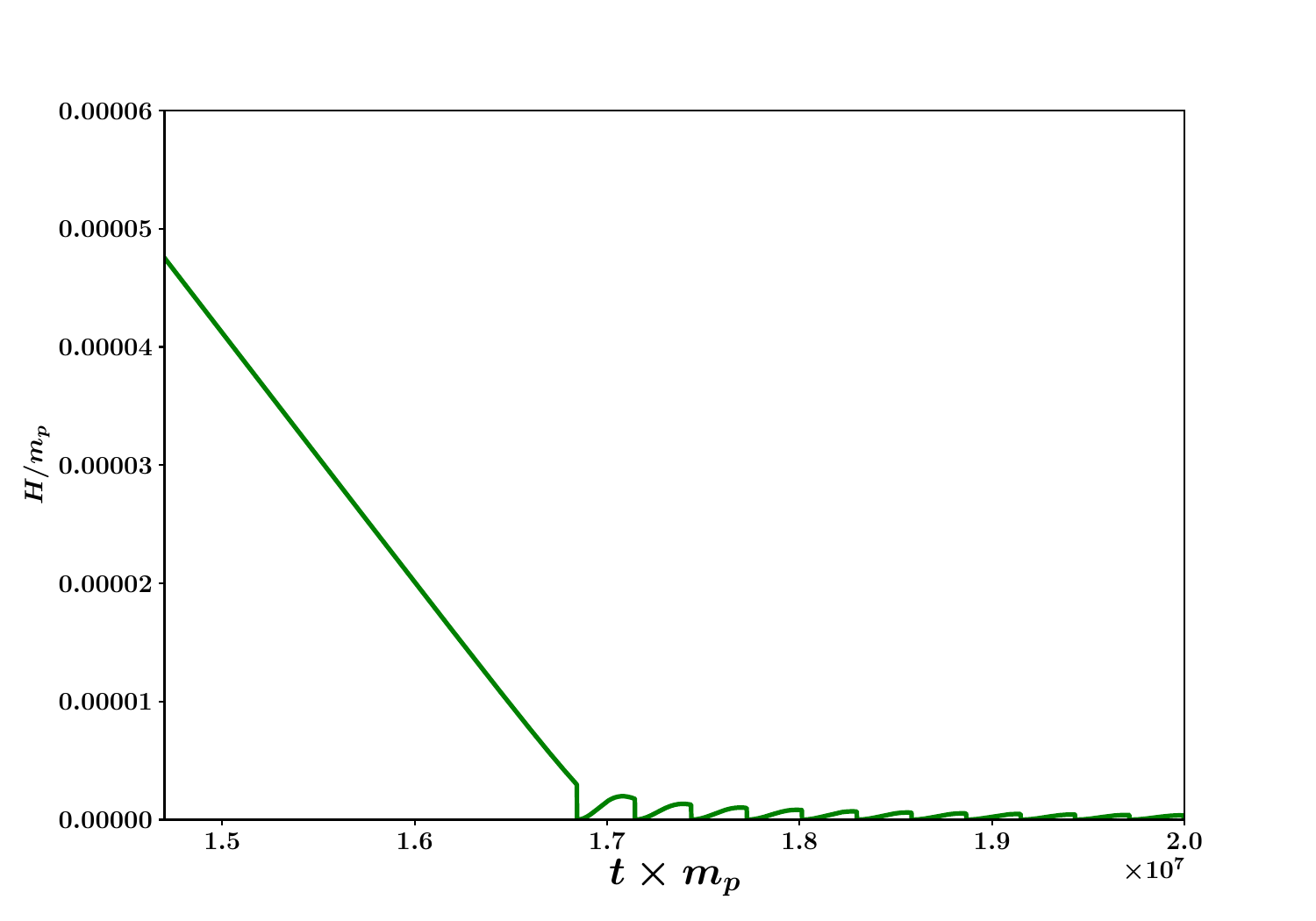}}
\subfigure[]{
\includegraphics[width=0.48\textwidth]{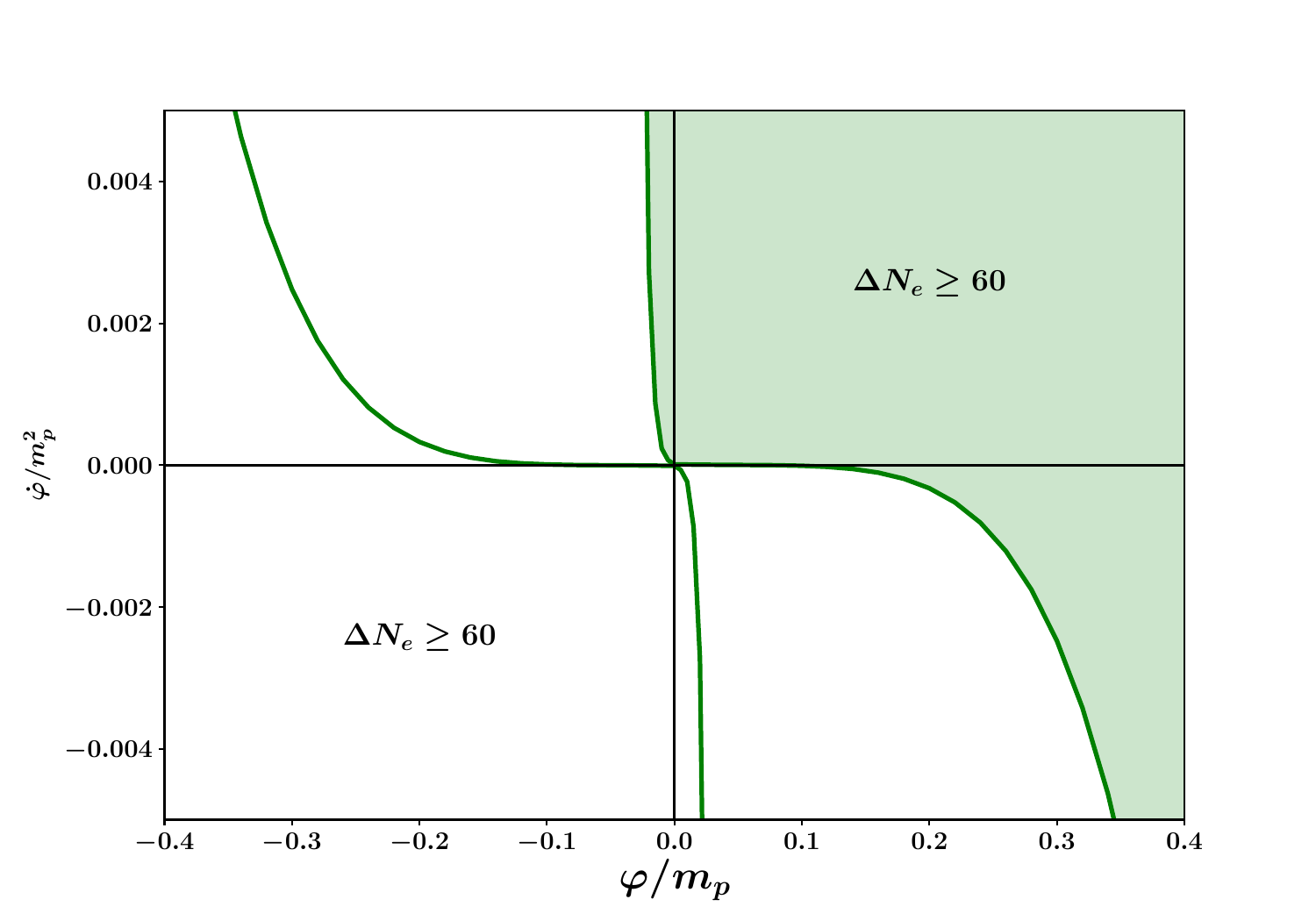}}
\captionsetup{
	justification=raggedright,
	singlelinecheck=false
}
\caption{{\bf a)} shows a typical inflationary trajectory in the Jordan frame of Higgs inflation (\ref{eqn:action_higgs}). Note that $H$ decreases linearly during the inflationary stage and starts to oscillate after the end of inflation. The initial conditions basin is shown in  {\bf b)} where the  black regions contain   initial conditions leading to  sufficient amount of Higgs inflation with at least 60 e-foldings in the Jordan frame.}
\label{f.Higgs}
\end{figure}

\ber
\l(m_p^2+\xi\varphi^2\r)\l(2\dot{H}+3H^2\r)=-\f{1}{2}\dot{\varphi}^2+U(\varphi)-4\xi H\varphi\dot{\varphi}-2\xi\l(\dot{\varphi}^2+\varphi\ddot{\varphi}\r)~, \\
\ddot{\varphi}+3H\dot{\varphi}-\xi \varphi R + \f{dU(\varphi)}{d\varphi}=0~,
\eer
where $R=6\l(2H^2+\dot{H}\r)$. The expression for Hubble parameter is
\beq
H^2 = \f{1}{3m_p^2}\l[\f{1}{2}\dot{\varphi}^2+U(\varphi)-3\xi\l(2H\varphi\dot{\varphi}+H^2\varphi^2\r)\r]
\eeq
 In this case our results, represented in the right panel of  figure \ref{f.Higgs}, are qualitatively similar to  the Einstein frame results \footnote {There is some interesting behaviour near $\varphi=0$ and phase phase is slightly different from that in the Einstein frame. However this does not change the qualitative behaviour of the phase-space that we are concerned about in this work and we would like to revisit this behaviour in a future work.} in the right panel of figure \ref{f.swag}.

\section{Starobinsky inflation in the Jordan frame}
\label{sec:staro_jordan}

In this section we consider the problem of generality of Starobinsky inflation directly in  the Jordan frame. The field equations resulting from action (\ref{lagrangeana}) are the following
\begin{equation}
E_{ab}\equiv G_{ab}+\beta H_{\: ab}^{(1)}=0\,,
\label{eq.campo}
\end{equation}
where 
\begin{eqnarray}
 &  & G_{ab}=R_{ab}-\frac{1}{2}g_{ab}R\,,\nonumber\\
 &  & H_{ab}^{(1)}=-\frac{1}{2}g_{ab}R^{2}+2RR_{ab}+2g_{ab}\nabla^2 R-2R_{;ab}\,. \label{H1}
\end{eqnarray}
Let us emphasize that since the  Einstein space $R_{ab}=g_{ab}\Lambda$ is an exact solution of (\ref{eq.campo}), all vacuum solutions of GR are also exact solutions of the quadratic theory \eqref{lagrangeana}. 

We choose the following line element 
\[
ds^2=-dt^2+a(t)^2d\vec{x}^2, 
\] 
such that the scale factor is $a(t)$ and the Hubble factor $H=\dot{a}/a$. 

Note that time in  the Jordan frame $dt$ and  in the Einstein frame $d\tilde{t}$ are related as 
\begin{equation}
dt=e^{-\tilde{\phi}/2}d\tilde{t} \label{tempos}
\end{equation}

In the zero spatial curvature case this choice allows us to write the field equations as
\begin{eqnarray}
&&2\beta\ddot{H}H+6\beta\dot{H}H^2-\beta\dot{H}^2+\frac{1}{6}H^2=0\label{00}\\
&&-12\beta\ddot{H}H-18\beta\dot{H}H^2-2\beta\dddot{H}-9\beta\dot{H}^2-\frac{1}{3}\dot{H}-\frac{1}{2}H^2=0,\label{11}
\end{eqnarray}
the first of which is the constraint equation, while all the other field equations are identically satisfied. 

 Substitution of the expressions 
 \begin{equation}
 R=6\dot{H}+12H^2,\;\; \dot{R}=6\ddot{H}+24 H \dot{H}
 \end{equation}
 into the constraint equation \eqref{00} yields the expression for   $H$ to be  
 \begin{equation}
 H=\frac{-6\beta \dot{R}+\sqrt{36\beta^2\dot{R}^2+6\beta R^2(2\beta R+1)}}{6(2\beta R+1)}~,\label{valor de H}
 \end{equation}
which is always positive for $\beta>0$ and $R>-1/(2\beta)$. 

Before proceeding  further, we must make few important  remarks. Staying  in the  Jordan frame in the context of $f(R)$ theories, it is easy to see that the field equations given by \eqref{eq.campo} can be expressed as 
\begin{eqnarray*}
&&f^\prime \left[ R_{ab}-\frac{1}{2}Rg_{ab}\right]+\frac{1}{2}g_{ab}\left[Rf^\prime-f\right]-\nabla_a\nabla_bf^\prime+g_{ab}\square f^\prime=T^{\mbox{{\tiny M}}}_{ab}\\
&&f^\prime \left[ R_{ab}-\frac{1}{2}Rg_{ab}\right]=T^{\mbox{{\tiny eff}}}_{ab},\;\;\rightarrow T^{\mbox{{\tiny eff}}}_{ab}=T^{\mbox{{\tiny M}}}_{ab}-\frac{1}{2}g_{ab}\left[Rf^\prime-f\right]+\nabla_a\nabla_bf^\prime-g_{ab}\square f^\prime,
\end{eqnarray*}
where $T^{\mbox{{\tiny eff}}}_{ab}$ is the effective behaviour of matter fields acting as source plus contributions from the scalar field and $f=R+\beta R^2$ as mentioned earlier. From these equations it follows that $f^\prime$ behaves as the  inverse of gravitational constant. 
 Looking at expression (\ref{RC}) it is possible to see that there is a minimum value for the Ricci scalar given by $R=-1/(2\beta)$. When $R=-1/(2\beta)$, $f^\prime$ is zero and physically, the gravitational constant diverges and so  as all the perturbations. When $R<-1/(2\beta)$ gravity becomes repulsive, and the scalar field would  then have  a complicated dynamics, different from what is  given by (\ref{Einstein_frame}), the conformal factor between the Jordan-Einstein frames would then be given by $\tilde{g}_{ab}=f^\prime g_{ab}$ with $f^\prime<0$. 

We now briefly remind the reader that for inflation in the Jordan frame, unlike in the  Einstein frame, we do not have the usual quasi-de Sitter solution, rather the  Ruzmaikin solution \cite{ruzmaikina1970quadratic} which  for the zero spatial curvature case  is an
asymptotic $t\rightarrow\infty$ isotropic solution with scale factor growing as 
\begin{equation}
a(t)\propto\exp\left(-\frac{t^{2}}{72\beta}\right), \label{ruzmaikina}
\end{equation}
which gives a type of inflation 
\begin{equation}
H=H_0-\frac{t}{36\beta}, \label{RR}
\end{equation}
with slow roll condition satisfied for $\beta>>1$.

\begin{figure}[htb]
\begin{center}
\begin{tabular}{c c} 
 \includegraphics[width=0.98\textwidth]{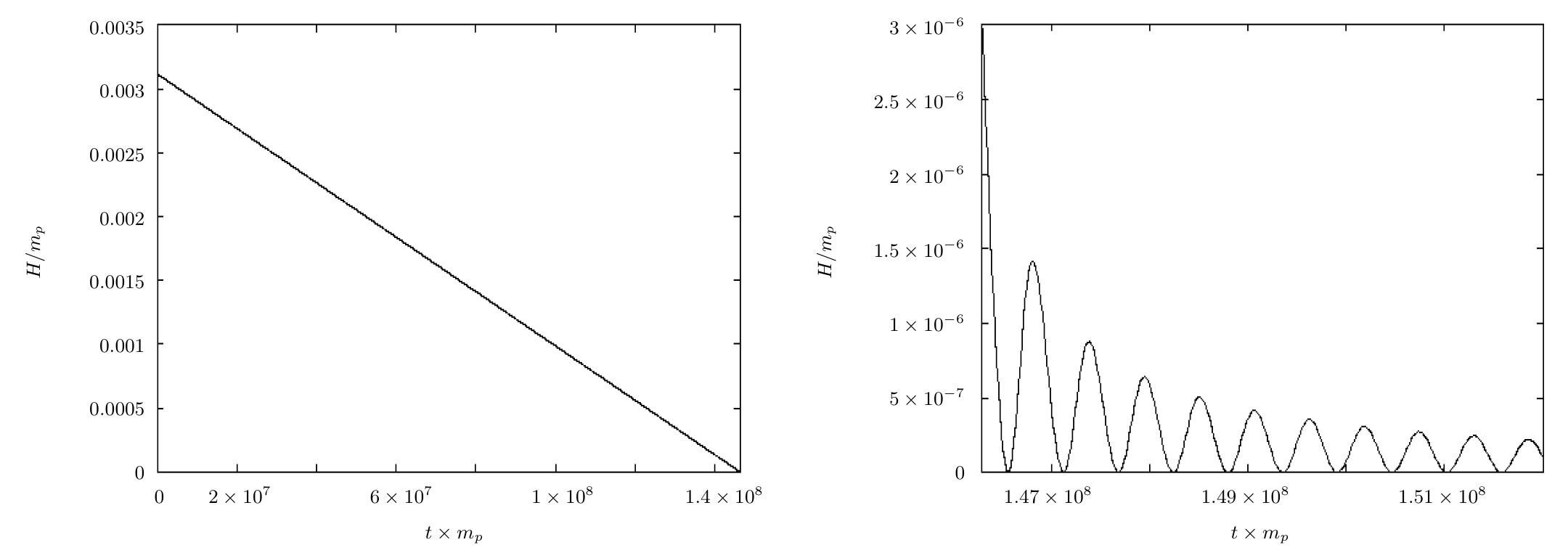} 
    \end{tabular}
\captionsetup{
	justification=raggedright,
	singlelinecheck=false
}
\caption{The time evolution of the Hubble parameter is shown for Starobinsky inflation in the Jordan frame. For an initial condition with  $\beta=1.305\times 10^{9}~m_p^{-2}$ and, $H=1.0\times 10^{-5} ~ m_p$ and $\dot{H}=1.1\times 10^{-6} ~ m_p^{2}$,  the left panel  {\bf a)} depicts Ruzmaikina Ruzmaikin solution \eqref{RR}, while the right panel {\bf b)} shows the  oscillatory regime after the end of inflation. The units are exactly the same as in the previous plot. }
 \label{figura1}
\end{center}
\end{figure}

This solution is stable into the future. In this context a problem remains as how to halt inflation when $\beta<0$ . Also, when $\beta<0$ there is the tachyon. Otherwise, when $\beta>0$, the Hubble parameter linearly decreases $H\rightarrow 0$, and later oscillates  as shown in figure \ref{figura1}.

 Our numerical results for the  generality of inflation have been  presented in figure \ref{f1} where  the region of initial conditions leading to sufficient amount of  inflation has been marked in black. The left panel depicts the basin of initial conditions  in  the Jordan frame
in terms of  variables $(H,R)$. For comparison, the right panel represents the corresponding diagram for the  Einstein frame in terms of variables $(\phi, \dot \phi)$.
Note that the boundary between initial conditions good for sufficient  inflation (in black) and bad for  sufficient inflation (white) is qualitatively
the same as that in the case of Higgs inflation in the Einstein frame as shown in the right panel of figure \ref{f.swag}. The only difference is that a  second ``good" region  for inflation, present in the right panel of figure \ref{f.swag} is
absent in the right panel of figure \ref{f1}  due to the presence of the steep left branch of the effective potential (\ref{potential}) shown in figure \ref{fig:pot_star}  and hence the whole diagram is not symmetric with respect
to changing the sign of $\phi$ and $\dot \phi$.

From  figure \ref{f1},  one can also notice  the agreement between the Einstein and  the Jordan frames. Particularly, an  initial negative
$\phi$ in the Einstein frame leads to sufficient inflation in the right panel. In the left panel, for the Jordan frame,  the same negative initial $\phi$ corresponds to negative initial $R$, so the inflationary attractor extends itself to negative $R$ in the Jordan frame. 
However, a particular 
feature of the left panel plot seems to be contradictory to the picture in the Einstein frame. Namely, according
to the left panel, one does not get sufficient amount of inflation starting from $R=0$ independent of  the initial value of $H$. On the contrary,
we know that inflationary trajectories, with adequate number of e-foldings, exist for initial $\phi=0$. Since zero $\phi$ corresponds 
to zero $R$, there seems to be an inconsistency. The goal of the next section is to resolve this  apparent paradox.

\section{Starobinsky Inflation: Comparison between the Jordan  and the Einstein  frame}
\label{sec:staro_jordanvseinstein}

We  next turn to obtain the map of initial conditions in the Einstein frame to the corresponding initial conditions in the Jordan frame. 

The expression for the Ricci scalar as well as   equation \eqref{RC} and its time derivative  taken together with the constraint \eqref{00} give rise to 
\begin{eqnarray}
&&R=\frac{e^{\tilde{\phi}}-1}{2\beta}\rightarrow \dot{R}=\frac{\dot{\tilde{\phi}}e^{\tilde{\phi}}}{2\beta} \label{RdotR}\\
&&H=-\frac{1}{2}\dot{\tilde{\phi}}+\frac{\sqrt{6}e^{-\tilde{\phi}}}{12}\sqrt{\frac{e^{\tilde{\phi}}(6\dot{\tilde{\phi}}^2\beta e^{\tilde{\phi}}+e^{2\tilde{\phi}}-2e^{\tilde{\phi}}+1)}{\beta}}\label{Hi}\\
&&\dot{H}=\frac{e^{\tilde{\phi}}-1}{12\beta} -2H^2 \label{DHi}\\
&&\ddot{H}=\frac{\dot{\tilde{\phi}}e^{\tilde{\phi}}}{12\beta}-4H\dot{H}~. \label{DDHi}
\end{eqnarray}
In the above mapping we have  picked up the positive root $H>0$ which satisfies the constraint \eqref{00}. Note  that the expressions for  higher derivatives in \eqref{DHi} and \eqref{DDHi} follow from the  definition of $H$ and $R$ in \eqref{RdotR} and \eqref{Hi}. While  studying the generality of inflation, we either plot the initial conditions space in the $(R,H)$ plane directly or by specifying the initial values  of  $(\tilde{\phi},\dot{\tilde{\phi}})$ and then  inserting into the expressions  \eqref{RdotR} and \eqref{Hi}.

\begin{figure}[htb]
\begin{center}
\includegraphics[width=0.98\textwidth]{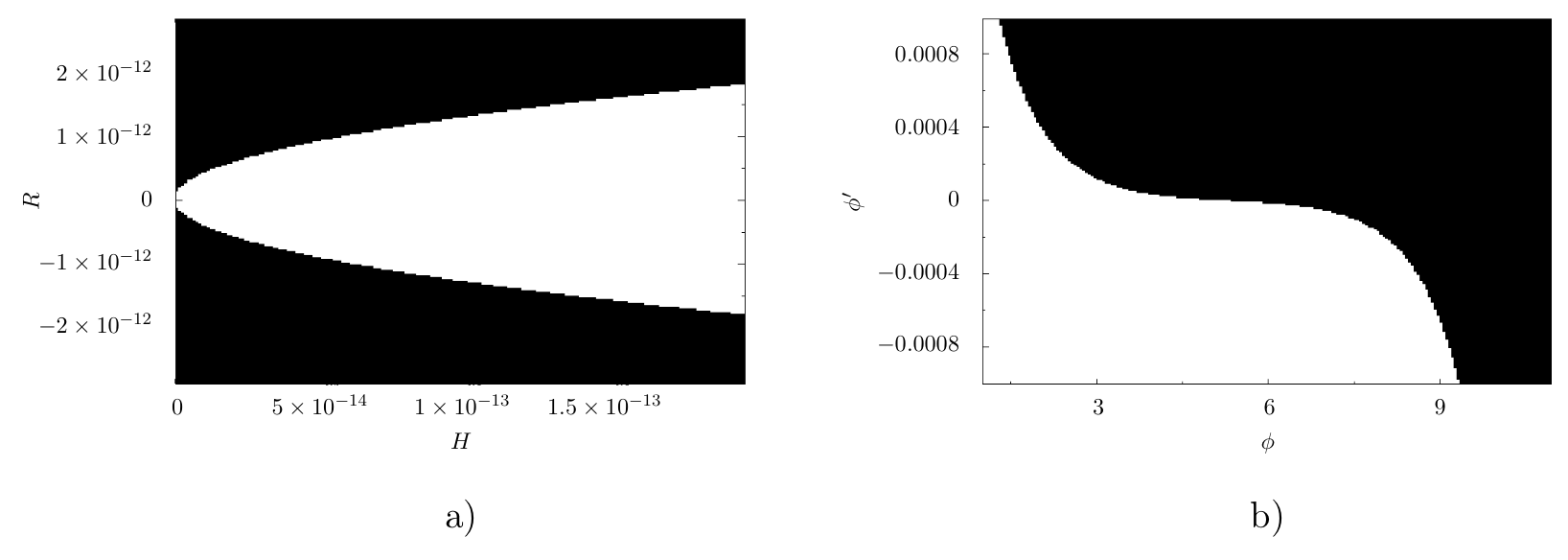}
\captionsetup{
	justification=raggedright,
	singlelinecheck=false
}
\caption{Initial conditions leading to sufficient inflation for Starobinsky inflation has been illustrated by the  black color region. The left panel {\bf a)} depicts the initial conditions  basin  directly in terms of  ($H$,$R$) coordinates in the Jordan frame. While the right panel {\bf b)} represents the same  in the Einstein frame where  we have chosen the variables $\phi$, $\phi^\prime$ with  $\phi$ in units of $m_p$ and $\phi^\prime$ in units of $m_p^2$ given by \eqref{phi-phi_tilda}. The  value of $\beta$ is chosen to be  $\beta=1.305\times 10^{9}$ in units of inverse square Planck mass, as  required by CMB observations.}
  \label{f1}
\end{center}
\end{figure}

There is a indeterminacy in the value  $H$ after fixing the $(\tilde{\phi},\dot{\tilde{\phi}})$ pair. Focusing on equation \eqref{Hi}, it can be noticed that initial $\tilde{\phi}=0$ with  $\dot{\tilde{\phi}}>0$ gives rise to vanishing initial value for $H=0$. This can be more clearly seen from the equation connecting Hubble parameter in both frames 
\begin{equation}
\tilde{H}=\frac{\tilde{a}^\prime}{\tilde{a}}=e^{-\tilde{\phi}/2}\left( \frac{\dot{a}}{a}+\frac{\dot{\tilde{\phi}}}{2}\right)=\frac{\dot{\tilde{\phi}}/2+H}{ \sqrt{2\beta R+1}}~,\label{TC---EF}
\end{equation}
where the prime denotes, according to (\ref{tempos}), 
\[
\tilde{\phi}^\prime=\frac{d}{d\tilde{t}}\tilde{\phi}=e^{-\tilde{\phi}/2}\frac{d}{dt}\tilde{\phi}=\frac{\dot{\tilde{\phi}}}{\sqrt{2\beta R+1}}~.
\]
From the expressions \eqref{TC---EF} and  \eqref{DHi}, we conclude that  the region of initial conditions with negligible value of the potential in the Einstein frame, for example near $\tilde{\phi}=0$ where $\tilde{H}\sim \dot{\tilde{\phi}}/2$, are mapped on to the origin of the $(H,R)$ plane in the Jordan frame.

The Jacobian of the map from $(\tilde{\phi},\dot{\tilde{\phi}})$ to $(H,R)$ region, as given in \eqref{RdotR} and \eqref{Hi},  is 
\begin{equation}
J=\frac{e^{\tilde{\phi}}\left(\sqrt{\Delta}-\sqrt{6}e^{\tilde{\phi}}\dot{\tilde{\phi}}\right)}{4\beta\sqrt{\Delta}}, \label{jacobiano}
\end{equation}
where 
\[
\Delta=\frac{e^{\tilde{\phi}}\left(6\dot{\tilde{\phi}}^2\beta e^{\tilde{\phi}}+e^{2\tilde{\phi}}-2e^{\tilde{\phi}}+1\right)}{\beta}.
\]
From \eqref{jacobiano} it can be seen that the determinant $J$ of the map   is zero for $\phi=0$ and $\dot{\phi}>0$, as expected from the above mentioned indeterminacy. Any initial value of $\dot{\tilde{\phi}}>0$ gets mapped  on to  the Jordan frame  initial conditions $H=0,\;\dot{H}=0,\;\ddot{H}=\dot{\tilde{\phi}}/(12\beta)$. While the dynamical equation \eqref{11} says that  the time evolution of this initial conditions with initial $\dddot{H}=0$ is well defined and it is non trivial. 
The set $(H,0)$ on  the $(H,R)$ plane,  corresponds to initial $\dot \phi <0$ and all such
trajectories are clearly non-inflationary.
This explains the apparent contradiction between numerical results of figure \ref{f1}.

\begin{figure}[htb]
\begin{center}
\includegraphics[width=0.75\textwidth]{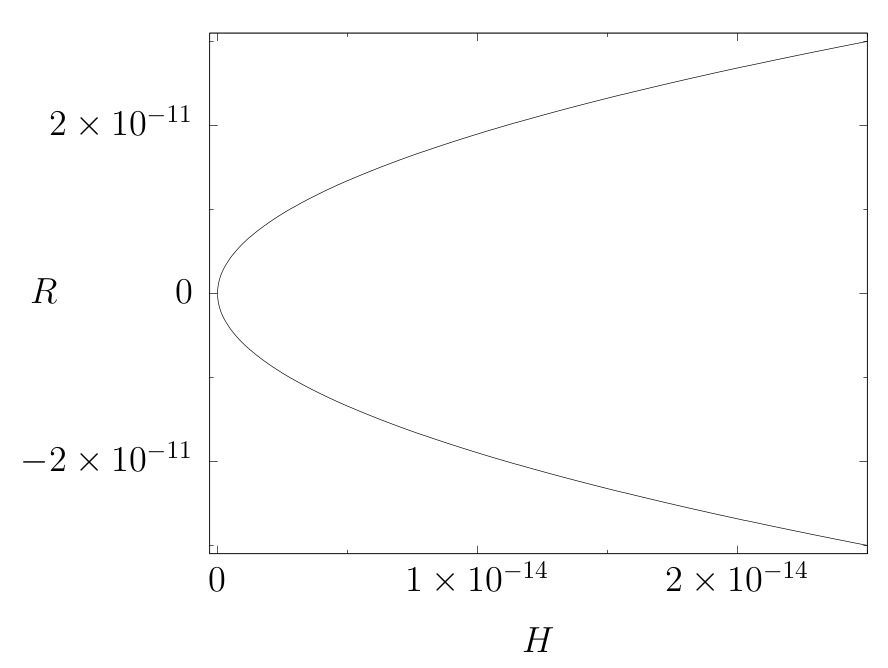}
\captionsetup{
	justification=raggedright,
	singlelinecheck=false
}
\caption{This figure is constructed with the same value of  $\beta=1.305\times 10^{9}$ as  used in figure \ref{f1}. The trajectory in the  ($H,~R$) space of  the Jordan frame  actually shows what happens when  the  Ricci scalar is negative initially. Any such trajectory would pass through the point $(0,0)$  in this plot.}
  \label{f2}
\end{center}
\end{figure}

In figure \ref{f2} we explicitly investigate the  trajectories in which  the Ricci scalar is negative initially. The conformal transformation between the  Einstein and the  Jordan frame only makes sense when $R>-1/(2\beta)$, which restricts the possible negative values of $R$ to a small region near the origin as shown in  the right panel of figure \ref{f1}. The trajectory  in  figure \ref{f2} in the ($H,~R$) space begins with an initial negative $R$, passes  through $R=0$ and then grows  to positive values of $R$, as expected from the correspondence between the Jordan and the Einstein frames. Besides that, figure \ref{f2} shows that the basin represented in figure \ref{f1} is an invariant set, as expected.

 From  figures \ref{f.swag} and  \ref{f1} {\bf b)},  it is clear  that figure \ref{f.swag} has two disjoints sets while in figure \ref{f1} there is only one set.  The reason for this is the fact that the potential in the Einstein frame for the particular gravitational theory given by the action (\ref{lagrangeana}) is very steep for large negative values of  $\phi$ which does not support inflationary trajectory \cite{FOSTER}. While the potential for Higgs inflation is symmetric for $\phi\leftrightarrow -\phi$ \cite{Mishra:2018dtg} so that inflation can occur for both positive and negative values of $\phi$. On the other hand, it is also possible to see that the upper right part of figure \ref{f.swag} {\bf b)} is qualitatively reproduced in figure \ref{f1} {\bf b)} as it should be, since both potentials (\ref{potential}) and (\ref{pot}) have the same behaviour for positive $\phi$.
 \section{Conclusions}
 \label{sec:conclusion}
 
 In the present paper we have considered the generality of Starobinsky inflation both in the Jordan as well as Einstein frame. The latter case has been considered in \cite{Mishra:2018dtg} for a rather standard set up where the initial energy of the effective scalar field is fixed and usually chosen to be  the Planck boundary.  It has been   argued in \cite{Gorbunov:2014ewa} that this choice, being physically motivated for a fundamental scalar field, may  not be very natural for an effective field. That is why we have lifted this requirement and analysed  the two dimensional set of initial conditions in the phase space $(\phi, \dot \phi)$. Due to similarity of potentials for Starobinsky and Higgs inflation we obtain results for
 generality of Higgs inflation as well. By giving up fixing the initial energy, have allowed ourselves to explain the nature of differences in the space of initial conditions leading to sufficient amount of  inflation for massive and plateau potentials, as remarked in \cite{Mishra:2018dtg}. The most striking example of this difference is the fact that initial field values close to $\phi=0$  can lead to sufficient inflation for a plateau potential, which is not the case for a massive scalar field potential.  We show that this difference originates from the physical requirement of the initial energy, not from any mathematical properties of the equations of motion. When we allow initial energy  not to be fixed, the resulting diagrams of  initial conditions yielding  sufficient inflation have  similar form  for both plateau and massive potentials.
 
 We have also constructed diagrams of initial conditions leading to sufficient Starobinsky inflation
 in the Jordan frame, using the variables $(H,R)$.
 The mapping $(\phi, \dot \phi) \to (H, R)$
 appears to be singular for $\phi=0$, which leads to inaccessibility of inflation starting from  initial $R=0$ despite the fact that $R=0$ corresponds to $\phi=0$ in the Einstein frame where sufficient inflation is possible starting from $\phi=0$. This results in completely different shapes of the phase space of  appropriate initial conditions for Starobinsky inflation in the $(H,R)$ plane (Jordan frame) as compared to that in the $(\phi, \dot \phi)$ plane (Einstein frame). Note that this does not happen for Higgs inflation since the phase spaces of  initial conditions have similar shapes for both the Einstein and Jordan frames. 
 
 \section*{Acknowledgments}

The  work of A.T. was supported by the Russian Science Foundation (RSF)
grant 16-12-10401
and by the Russian Government Program of Competitive Growth of Kazan Federal University. A.T. also thanks IUCAA, where part of this research work was carried out,  for their hospitality. D. M. would like to thank the hospitality of the Sternberg Astronomical Institute were this work was finalized and the Brazilian agency FAPDF {\it visita t\'ecnica} no. 00193-00001537/2019-59 for partial support. S.S.M. thanks the Council of Scientific and Industrial Research (CSIR), India, for
financial support as senior research fellow.

\end{document}